\documentclass[%
 reprint,
 amsmath,amssymb,
 aps,
 showkeys,
]{revtex4-2}

\usepackage[dvipdfmx]{graphicx}
\usepackage{dcolumn}
\usepackage{bm}

\usepackage{color}
\definecolor{Red}  {rgb}{1,0,0}
\definecolor{Green}{rgb}{0,1,0}
\definecolor{Blue} {rgb}{0,0,1}
\usepackage{bm}
\newcommand {\bfv}[1] {{\boldsymbol {#1}}}

\newcommand {\IND} {\hspace*{10pt}}
\newcommand\Rey{\mbox{{\rm Re}}}     
\newcommand {\com}[1] {}

\begin{document}

\preprint{APS/123-QED}

\title{Revisiting visualization of spiral states in a wide-gap spherical Couette flow}


\author{Isshin Arai}
\affiliation{
  Department of Pure and Applied Physics,
  Faculty of Engineering Science, Kansai University,  Osaka, 564-8680, Japan
}
\author{Tomoaki Itano}
\email{itano@kansai-u.ac.jp}
\affiliation{
  Department of Pure and Applied Physics,
  Faculty of Engineering Science, Kansai University,  Osaka, 564-8680, Japan
}
\author{Masako Sugihara-Seki}
\affiliation{
  Department of Pure and Applied Physics,
  Faculty of Engineering Science, Kansai University,  Osaka, 564-8680, Japan
}

\date{\today}

\begin{abstract}
  A pioneering study conducted by Egbers and Rath (1995) [Acta Mech. 111 pp. 125--140 (1995)] experimentally captured spiral waves { to elucidate the transition in the wide-gap spherical Couette flow}. However, the physical field quantities of the spiral waves corresponding to light patterns of various intensities, as obtained in the experiment, remain unclear, { and we have yet to move beyond the understanding that the reflected light from shear-sensitive flake tracers responds to a flow that appears at the transition}.
  In this study, the experiment to visualize spiral waves using aluminum flakes, as performed by Egbers and Rath, was numerically reproduced by solving the translational and rotational motions of the particles in a spiral wave.
  First, the spiral wave in a spherical Couette flow with an aspect ratio $\eta=1/2$ was numerically calculated using the Newton--Raphson method.
  Subsequently, the image that was numerically reproduced from the spiral wave was compared with an experimentally visualized image.
  The torque acting on the inner sphere and the phase angular velocity of the spiral waves with various wavenumbers were { provided}.
{ Attempts have been made to determine the instantaneous physical quantity to which the light and dark patterns obtained in the visualization corresponded, and the orientation motion of the flakes { developed} in the advective history of the flow is essential to yield favorable results.}
  { Exploring the correlation between flow visualization results and shear structures may provide a new avenue for quantitatively estimating spatial structures and time scales in complex and quickly time-varying flow fields, such as turbulence.
  }
\end{abstract}


\maketitle


\section{introduction}

\IND
The fluid phenomena associated with astrophysical bodies, such as those in their cores, atmospheres, and oceans, involve flows among concentric double spheres\cite{Fow04,God88}.
  Although considering the thermohydrodynamic or electromagnetic factors for the flow in the core of an astrophysical body is important\cite{Cha61,Bus75,Zeb83,Feu11,Kid97,Sak99}, even considering only the angular velocity difference among spherical boundaries complicates the flow.
  We considered a system involving the Newtonian fluid flow confined in the gap between spheres of radii $\tilde{r}_{\rm in}$ and $\tilde{r}_{\rm out}$, where the outer spherical boundary is at rest and the inner spherical boundary rotates at a constant angular velocity $\tilde{\Omega}_{\rm in}$.
  This system is referred to as the {\it spherical Couette flow } (SCF)\cite{Mun75,Bel84,Egb95,Nak02,Wic14,Bar18,Hof19}.

\IND 
The SCF is determined based on two dimensionless parameters, that is, the aspect ratio $\eta=\tilde{r}_{\rm in}/{\tilde r}_{\rm out}$ and Reynolds number $\Rey=\tilde{r}_{\rm in}^2 \tilde{\Omega}_{\rm in}/\tilde{\nu}$, where $\tilde{\nu}$ is the kinematic viscosity of the working fluids.
Egbers and Rath\cite{Egb95} obtained a phase diagram in the $\eta-\Rey$ parameter space based on the assumption that regardless of the initial conditions, these parameters only determine the final equilibrium state via nonlinearity and dissipation. 
{ Realized} below the critical Reynolds number determined for each value of $\eta$, the axisymmetric fundamental flow is not considered a laminar flow because the streamlines are not parallel to the spherical boundaries.
Both spherical boundaries must be non-slipping { to generate} the axisymmetric basic flow because { the rigidly rotating flow is realized} if the slip condition is imposed on either spherical boundary.
For small values of $\eta$ { ($\eta\lesssim 3/4$)}, also referred to as the ``wide gap case''\cite{Nak02}, the flow becomes unstable because of the crossing fundamental flow with inflection points from mid-latitudes to the poles above the finite critical Reynolds number, leading to a {\it spiral wave}, as described in \cite{Ara97,Hol06,Abb18a,Abb18b}.

\IND
{ The geomagnetic field on a planet is formed by convection currents in the liquid core. Considering the planets in the Solar System that form the geomagnetic field, the radius ratio of the inner core to the outer core varies from 0.35 for Earth to 0.55 for Mercury.
} In this study, we focus on a wide spherical SCF { with} a radius ratio $\eta=1/2$.
Belyaef et al. \cite{Bel84,Bel91} estimated the first transitional Reynolds number for $\eta=1/2$ using power spectra obtained from laser Doppler velocimetry.
A recent study \cite{Dum94,Jun00} showed that the basic flow becomes unstable against infinitesimal sinusoidal disturbances for an azimuthal wavenumber of 4 at $\Rey=489$.
  Immediately after $\Rey$ exceeds the critical Reynolds number at which a spiral wave with a wavenumber of 4 is expected to be sustained, a spiral wave with a lower wavenumber, such as 3 or 2, appears dominant as the Reynolds number increases, which has been observed in experiments\cite{Bel78,Yos23}.
  The classical visualization technique can visualize the flow state in a spiral wave as a spiral pattern\cite{Egb95,Wul99} with $m$ equally spaced arms extending from the poles to the equatorial zone in each hemisphere.
This phenomenon is also observed { as a polygonal pattern} in the flow on a rotating planar disk in a stationary casing\cite{Miz09}.
Belyaef et al. \cite{Bel84,Bel91} reported hysteresis loops related to spiral waves with various wavenumbers in multiple ranges of $\Rey$.
The numerical experiments based on the shooting method indicate that both spiral waves with wavenumbers 3 and 4 possess independent basin attractions in the phase space\cite{Got21}.
However, to date, the reasons why experiments predominantly produce spiral waves with smaller wavenumber remain unclear.

\IND
{ In the 1970s, when SCF research began to flourish, the transitions in SCF were measured by visualization technique by the addition of aluminum to the fluid\cite{Bel78}.}
We revisit the pioneering study by Egbers and Rath (1995), wherein spiral waves were experimentally captured.
They suspended a small amount of aluminum flakes in a working fluid at a concentration of 0.05\% as tracer particles to visualize the flow structure. In their experiment, the viscosities of the working fluids with tracer particles were undetectable in rheometer measurements.
They successfully captured the transition of the visualization image on the path from the basic state to the spiral wave by setting a light source and video camera through a half mirror in the polar direction of the SCF experiment.
The change in the captured pattern was not due to a spatial fluctuation in flake density.
This is because, as long as the particles that are neutrally suspended in the fluid are sufficiently small, the number density of the particles in the flow must be uniform, regardless of whether they are spherical or flake-shaped.
However, in principle, flakes rotate under the shear of the flow such that the flow structure as a vorticity is { educed} due to the orientation.

\IND
However, the physical field quantities of the spiral wave to which the bright and dark patterns correspond remain unclear.
{ When light is passed through the suspension with flakes, the intensity of the scattered light is very sensitive to the orientation of the flakes, indicating that even small velocity gradients can be measured\cite{Bel78}. Thus, flow velocity gradients may be measured from light intensity using visualization of flake-mixed fluid experiments; thus, can we conclude that the spatial pattern of light intensity is solely determined by the instantaneous velocity gradient?}
In this study, we virtually distributed infinitesimal flake particles with a uniform density in an exact field of spiral waves, which were numerically solved using spectral expansion and the Newton--Raphson method\cite{Got21}.
By solving the translational and rotational motions of the particles in the spiral wave using the procedure presented by Goto et al. \cite{Got11}, we virtually reproduced the light and dark patterns of the flow field photographed in the experiment by Egbers and Rath\cite{Egb95}. Furthermore, we { discerned} the physical quantities to which the pattern of brightness corresponds.

The remainder of this paper is organized as follows:
The next section describes the numerical method for solving spiral waves via the Newton--Raphson method.
To visualize the spiral state, the equations for both translational and rotational motions of an infinitesimal drifting planar particle are presented in a flow and numerical procedure.
In Section 3, we compare the images of the spiral state visualized experimentally in \cite{Egb95} and numerically using our approach.
In Section 4, we discuss the physical quantities obtained from images using the classical visualization technique.
Finally, Section 5 summarizes this study.

\section{Numerical Method}

Under the nondimensionalization based on the half length of the gap width, $\Delta \tilde{r}$, and viscosity, we assume that an inner sphere centered at the origin rotates at a constant angular velocity $\Omega_{\rm in}=\frac{\tilde{\nu}}{\Delta \tilde{r}^2}\tilde{\Omega}_{\rm in}$ about the z-axis, and we consider the polar coordinates $(r,\theta,\phi)$.
The nondimensionalized momentum equation of an incompressible Newtonian fluid confined in the gap $2\Delta \tilde{r}$ between the inner and outer spheres is given by 
\[ \frac{\partial\bfv{u}}{\partial t}+\bfv{u}\cdot\bfv{\nabla}\bfv{u}=-\bfv{\nabla}p+\bfv{\nabla}^2\bfv{u}\]
under nonslip conditions on both the inner and outer spherical boundaries.
The arbitrary incompressible velocity field $\bfv{u}$ can be expressed as $\bfv{\nabla}\times\bigl(-\bfv{r}\varPsi + \bfv{\nabla}\times(\bfv{r}\varPhi)\bigr)$ according to two scalar fields, $\varPsi(r,\theta,\phi)$ and $\varPhi(r,\theta,\phi)$, { commonly known as} toroidal--poloidal decomposition\cite{Bel78}.
Each scalar field may be described according to the product of a series of spherical harmonics in $\theta,\phi$ directions and modified Chebyshev polynomials in the radial direction.
If the solution is a rotating wave solution with constant phase angular velocity $\omega$, the time-derivative term in the governing equation $\frac{\partial}{\partial t}$ can be replaced by $-\omega\frac{\partial}{\partial \phi}$.
Thus, all the terms in the governing equations of $\varPsi$ and $\varPhi$ specifying a spiral state can be numerically evaluated.
The condition satisfied by the toroidal and poloidal components of a spiral wave is equivalent to an algebraic equation with degrees of freedom representing its flow state.
The algebraic equation was solved using the Newton–-Raphson method with the aid of open numerical libraries\cite{Sch13,Fri05,And99}.

\IND
{
  When the spiral wave (or secondary wave) was first experimentally discovered in the pioneering works, it was captured as a periodic light reflection and was perceived as a wave of disturbance with wavenumber $m$ that could be superimposed.
  On the other hand, it is a solution with the $m$-fold symmetry to the Navier--Stokes equation for us, which cannot be superimposed, and its amplitude is uniquely determined by $\Rey$ and $\eta$; thus, we refer to it as {\it spiral state} rather than as a spiral wave.
}

\IND We considered two arbitrary material lines $\bfv{a}$ and $\bfv{b}$ with infinitesimal length floating at a point in the spiral state.
In the flow, the location, length, and direction change with time $t$.
The time evolutions of $\bfv{a}(t)$ and $\bfv{b}(t)$ are given by
$\dot{\bfv{a}}=\bfv{a}\cdot\bfv{\nabla}\bfv{u} + {\rm O}(|\bfv{a}|^2)$,
where $\bfv{u}$ is the velocity field at $\bfv{r}(t)${, which is the position of the material line at time $t$\cite{Bat67}.}
The positions of the material lines follow the order of $\dot{\bfv{r}}=\bfv{u}(\bfv{r}(t),t)$.
Owing to the flow incompressibility, the time variation of the infinitesimal surface element formed by these material lines $\bfv{S}(t)=\bfv{a}\times\bfv{b}$ satisfies the equation $\dot{\bfv{S}}=-\bfv{
\nabla}\bfv{u}\cdot\bfv{S}$.
However, the areas of the small and thin flake-shaped particles distributed in the flow do not change over time.
The unit normal vector of the flake $\bfv{s}$ is represented by $\bfv{s}(t)=\bfv{S}/|\bfv{S}|$ and $d|\bfv{S}|^2/dt=-2\bfv{S}\cdot\bfv{\nabla}\bfv{u}\cdot\bfv{S}$. 
Therefore, the orientation of the flake at time $t$ is given by the time integration of the equation: $\dot{\bfv{s}}=\bfv{s}\times\bfv{s}\times\bfv{\nabla}(\bfv{u}\cdot\bfv{s})$, as shown in \cite{Got11}.
We obtained the orientation of the flakes drifting in the spiral state of the SCF by numerically solving the equation \cite{Yos23}.

\IND For illustration, we consider a flake located in a simple shear flow $\bfv{u}=y\bfv{e}_x$, where the flake moving at a constant speed along the x-axis changes its orientation with time.
The components of $\bfv{s}$ are denoted as $(s_x,s_y,s_z)$, and the time evolution of each component is described according to the ordinary differential equations $\dot{s_x}=s_x^2s_y$, $\dot{s_y}=s_x(s_y^2-1)$, and $\dot{s_z}=s_xs_ys_z$.
The system possesses two conserved quantities: $s_x^2+s_y^2+s_z^2$ and $s_x/s_z$.
The steady solution is $s_x=0$.
Under arbitrary initial conditions, $\bfv{s}$ moves away from $s_z=\pm 1$ and asymptotically approaches $s_y=\pm 1$ over time.
Thus, the attractors, repellers, and limit cycles may exist in the phase space of the flake orientations (unit spherical surface) for an arbitrary steady flow.
For general time-dependent flow fields, invariant sets solved in an instantaneous flow field may vary with time.
This example indicates that even if the orientation of the flakes initially scattered at a point in the flow is stochastically isotropic, they may align in a specific direction after a certain time.
Conversely, the orientations of the flakes concentrated in a specific direction may become isotropic after a certain period.

As incident collimated rays from a laser or other light sources { outside the container} pass through the fluid in a spherical Couette flow container, some rays may be reflected by the flakes.
When a ray reflected by the flakes enters the camera, the image is brighter at the position of the flakes.
Based on the assumption of no light attenuation in the fluid, the condition for the flake to completely reflect a ray from the light source to the camera was evaluated.
The direction of the incident ray on the flake is represented as a unit vector $\bfv{I}$, which is independent of the position of the flake.
The reflected light is observable if the orientation of flake $\bfv{s}$ is in the bisectional direction between $-\bfv{I}$ and the optical axis of the camera.
Generally, the direction of a ray $\bfv{R}$ reflected from the flake is $\bfv{R}=\bfv{I}-2(\bfv{I}\cdot\bfv{s})\bfv{s}$. The alignment of the reflected rays against the optical axis of the camera must be 
correlated with the intensity distribution in the recorded images.
Thus, the intensity of the reflected light captured by the camera may be evaluated using the magnitude of $\bfv{C}\cdot\bfv{R}$, where $\bfv{C}$ is a unit vector along the optical axis of the camera lens with respect to the reflecting flake.
If the flakes are isotropically oriented in a localized region, the light intensities are scattered throughout the region such that the region has{ a gritty pattern}.
{ On the other hand,} if the flakes are aligned in a direction in the localized region, { the light intensities captured by a camera in the region remain consistent regardless of the flakes reflecting light} such that the region may appear with a smooth pattern.

\section{Results}
  \begin{figure}[ht]
    \centering
    \includegraphics[angle=0,width=0.90\columnwidth]{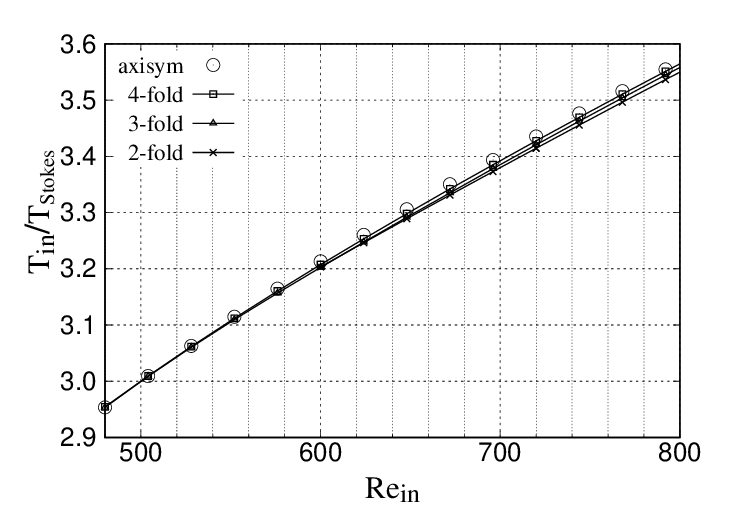}
    \caption{The normalized value of the torque acting on the spherical boundary against the Reynolds number,
   which shows a bifurcation diagram of the axisymmetric and 2-, 3-, and 4-fold states.
    }
    \label{fig:torque}    
    \end{figure}
  For $\eta=0.5$, the 4-fold state bifurcates at $\Rey\sim 489$ from the axisymmetric state, and a 3-fold state occurs at $491$\cite{Got21}, followed by the 2-fold state bifurcating { around $\Rey\sim 600$.}
  All states originated in the axisymmetric state, which first lost stability at $\Rey=489$ against a perturbation with a wavenumber of 4.
  The torque values exceeded the Stokes flow by a factor of 3 \cite{Lan87} in the range of $\Rey>500$.
  The torque values of the $m$-fold states are smaller than those of the axisymmetric state, and the torque slightly decreases as the wavenumber $m$ decreases.
  This may be attributed to the fact that a spiral state with a smaller wavenumber is predominantly obtained in experiments at a high $\Rey$.
  This is contrasting to previous experimental results on narrow-gap SCF\cite{Nak78}, { where the frictional torque of states successively bifurcated from the axisymmetric state is larger than that of the predecessor, such as the axisymmetric state}, as measured in a turbulent transition.
  Intuition indicates that the transition from laminar to turbulent flow becomes more complex (as the wave number increases) with increasing Re. Simultaneously, it is accompanied by a greater dissipation and friction.
 However, the spiral state bifurcation in SCF suggests that this intuition is not always correct.

The phase angular velocities of the spiral states calculated using the Newton--Raphson method have similar values regardless of the wavenumber and are found to be in the range of approximately $1/6\le \omega/\Omega_{\rm in}\le 1/7$.
As with the torque graphs, confirming the difference among the branches with different wavenumbers in the graphs indicating the relationship between Re and phase speed is difficult. In addition, the value of the dimensionless frequency proposed by Belyaef\cite{Bel84} may be utilized to better represent the difference.
This quantity is the frequency of the flickering of the light reflected from the flow, which can be experimentally measured (normalized by the rotation rate of the inner sphere).

  \begin{figure}[ht]
    \centering
    \includegraphics[angle=0,width=0.98\columnwidth]{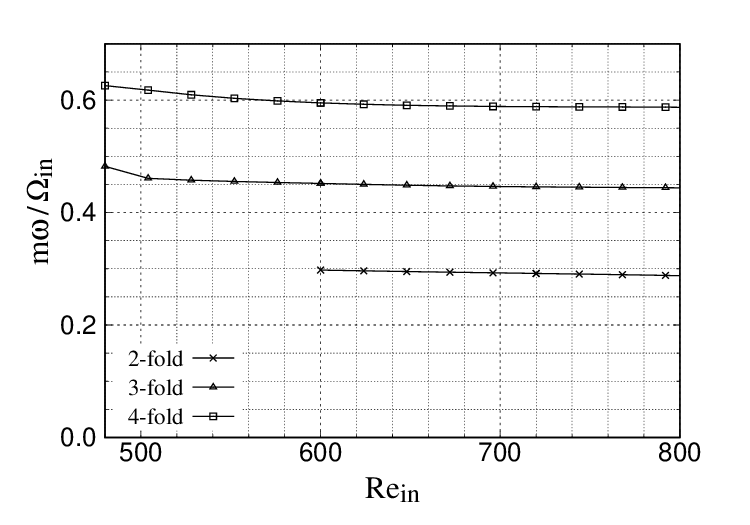}
    \caption{
     Two-, three- and four-fold states and branches using dimensionless frequency
    }
  \end{figure}

      \begin{figure}[ht]
      \centering
      \includegraphics[angle=0,width=0.78\columnwidth]{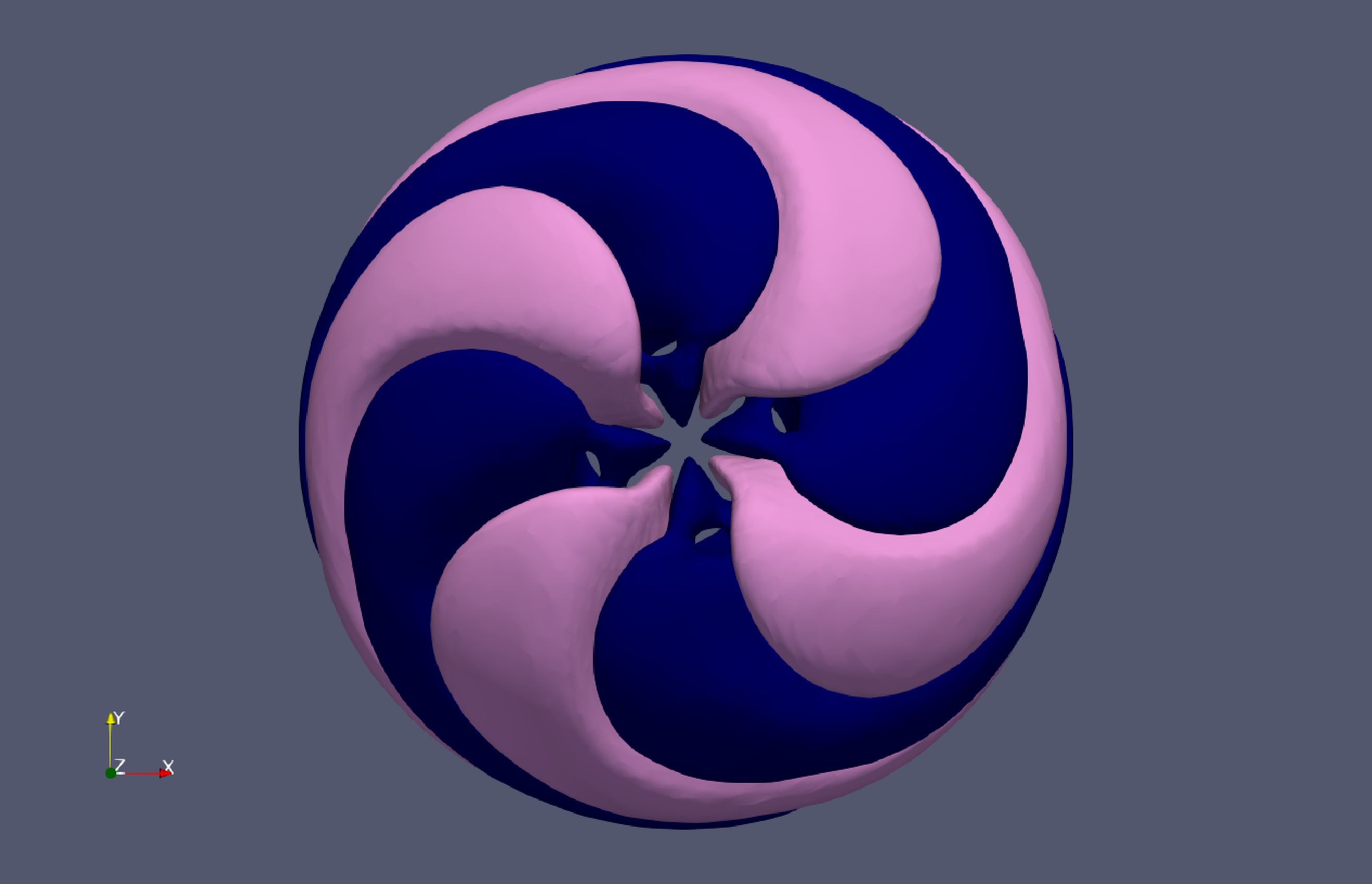}
      \caption{
        The isosurface of $u_r$ and $\varPhi$ of the 4-fold state at $\Rey=600$ was captured from $+z$ direction. The blue and pink isosurfaces correspond to the region with $u_r>1$ and $\varPhi > 0.1$, respectively.
      }
      \label{fig:polar view}
    \end{figure}

\begin{figure}[t]
      \centering
      \includegraphics[angle=0,width=0.70\columnwidth]{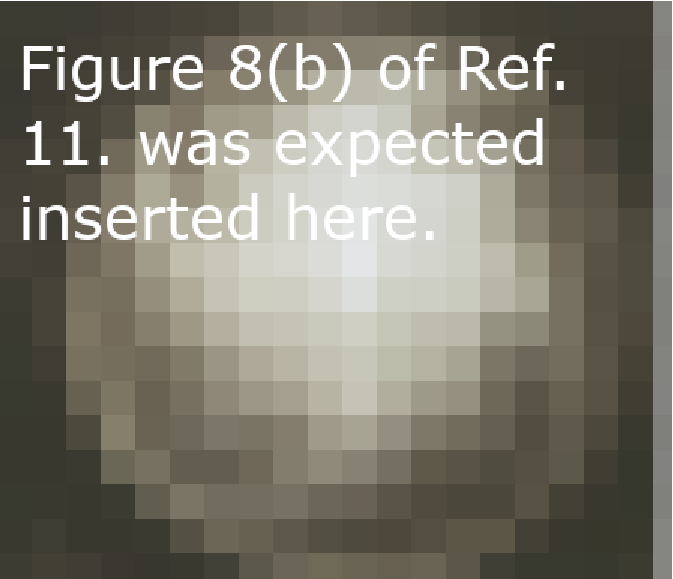}
      \caption{
        Figure expected to be inserted here (Figure 8(b) of Ref.11) is an experimental visual observation at $\Rey=1583$ using both a white light source and video camera via a half mirror positioned at $+z$.
        The captured image was identified as ``non-axisymmetric secondary wave'' with four spiral arms for $\eta=2/3$.
        Suspended aluminum flakes advected in the working fluid are employed as tracer particles
      }
     \label{fig:Egb}
    \end{figure}

Fig.\ref{fig:polar view} illustrates the overall view of the 4-fold spiral state taken from the $+z$ direction, where the inner spherical boundary rotates anticlockwise.
The blue isosurface corresponds to the region where $u_r>1$ and pink isosurface corresponds to the region where $\varPhi>0.1$.
Both images imply the term ``spiral,'' which describes the flow structure consisting of equally spaced arms extending from the poles to the equatorial zone in each hemisphere.
Through an experimental visualization technique using aluminum flakes drifting through a horizontal plane illuminated by a laser sheet, Yoshikawa et al.\cite{Yos23} recorded a spiral wave with a wavenumber of 3 on the horizontal cross-section over the critical Reynolds number.
By solving the equation of motion for infinitesimal planar particles advecting in the flow field of a spiral wave, they also virtually reproduced a sequence of visual distributions of reflected light. This is in good agreement with the pictures recorded experimentally.
We applied the same method to reproduce an image of the 4-fold spiral state (Fig.\ref{fig:Egb}) captured by Egbers and Rath in \cite{Egb95}, which was the pioneering observation to capture the transition from the axisymmetric state in a wide-gap case.

Fig.\ref{fig:Egb} is a copy of Fig.8 in \cite{Egb95}, where the 4-fold spiral state, illuminated by a light source from the direction of a pole, was captured using a camera through a half mirror placed at the pole.
The incident rays from the polar direction are partly absorbed in the bulk fluid, partly reflected by the surfaces of the flakes floating in the bulk fluid, and partly transmitted through the inner fluid layer.
By simplifying the aforementioned complex process followed by the light rays, we consider the image of the 4-fold spiral state captured by the camera as a superposition of the rays reflected toward the optical axis of the camera by the flakes existing on various $z$ planes.

    \begin{figure}[ht]
      \centering
      \includegraphics[angle=0,width=0.78\columnwidth]{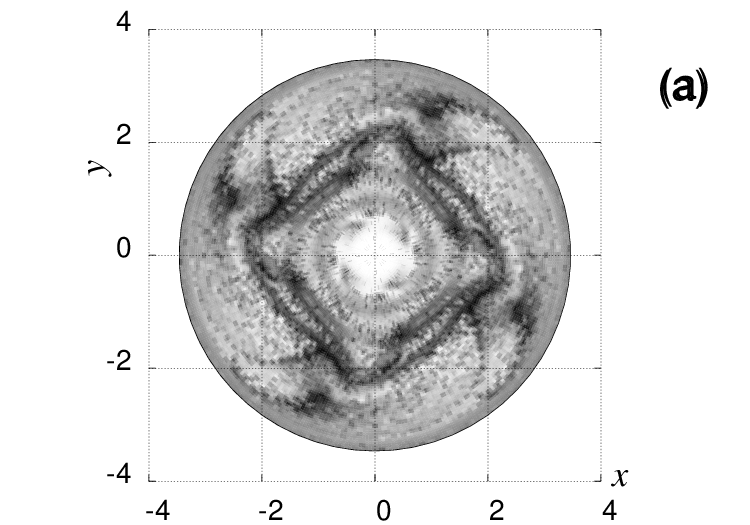}
      \includegraphics[angle=0,width=0.78\columnwidth]{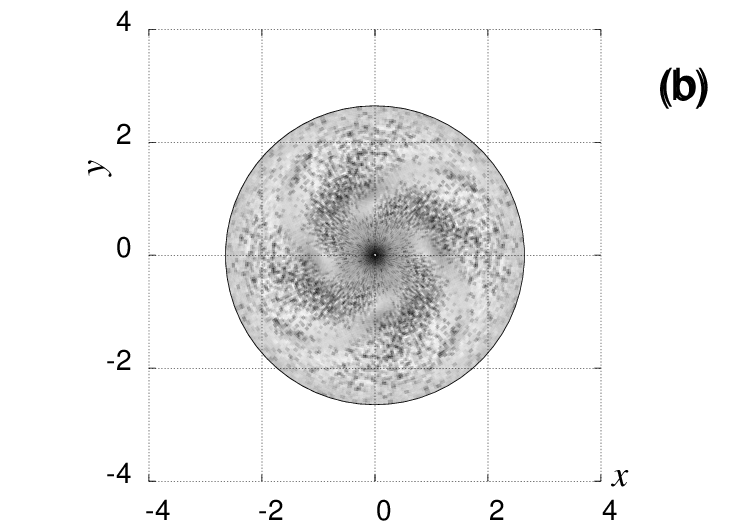}
      \includegraphics[angle=0,width=0.78\columnwidth]{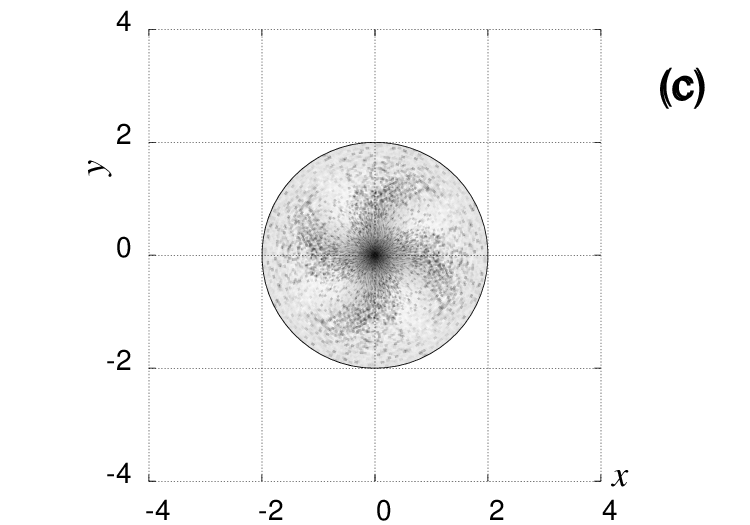}
      \caption{
The numerical visualization of the 4-fold state solved at $\Rey=600$ was virtually reproduced in the case of $\bfv{I}=-\bfv{e}_z$ and $\bfv{C}=+\bfv{e}_z$, which shows a camera and light source positioned at $+z$ direction. Inner products with values varying from 0 to 1 are illuminated in a black-to-white grayscale. All initial orientations of flakes are random with respect to space. From top to bottom $z=2,3,\sqrt{12}$.
      }
      \label{fig:virtual}
    \end{figure}

  The light intensity of the rays reflected by { $32000$ flakes} distributed at different values of $z$ was calculated for a 4-spiral state solved at $(\eta,\Rey)=(1/2,600)$ using the computational visualization method described in the previous section, where we assumed the direction of the incident ray and optical axis of the camera as $\bfv{I}=-\bfv{e}_z$ and $\bfv{C}=+\bfv{e}_z$, respectively.
  Figs.\ref{fig:virtual}(a), (b) and (c) correspond to the light intensity reflected at $z=2, 3$, and $\sqrt{12}$, respectively.
{ Flakes were isotropically oriented using random numbers at $3 T_{\rm in}$ before the time to drift to each position on each $z$, where $T_{\rm in}=2\pi/\Omega_{\rm in}$.}
At the poles, there is a flow from the outer sphere to the inner sphere in the radial direction due to the Ekman pumping.
This flow is expected to incline the normal vector of the flake equatorward, except near the inner sphere.
The flakes move along the flow while satisfying $\bfv{u}\cdot\bfv{s}=0$ for relatively long sections, resulting in the appearance of a small round shadow near the pole in Figs.\ref{fig:virtual}(b) and (c).

Fig.\ref{fig:virtual}(a) shows that a square streaky shadow surrounding the inner sphere.
The square pattern appears to be related to the cross-sectional image of the shear arms extending from the poles in the spiral state, as shown in Fig.\ref{fig:polar view}.
However, the physical quantity of the state related to the shadow region cannot be easily explained, and this will be discussed in the next section.
The bright regions in both $z=3$ and $z=\sqrt{12}$, which exist in the foreground from the camera side, are superimposed on the square shadow in Fig.\ref{fig:virtual}(a). The square pattern is invisible to the camera located at the pole direction.
By contrast, the horn-shaped shadows extending outward from the four vertices of the square streaky shadow shown in Fig.\ref{fig:virtual}(a) do not disappear even if the bright images of $z=3$ and $z=\sqrt{12}$ are superimposed.
The four horn-shaped shadow patterns shown in Fig.\ref{fig:Egb} in \cite{Egb95} correspond to those in Fig.\ref{fig:virtual}(a).
This suggests that the inner sphere in \cite{Egb95} also rotates anticlockwise,
{ as same as in}
the case performed in this study.   
Moreover, the intensity of light at $z=\sqrt{12}$ is stronger than that at $z=3$. 
This suggests that the normal vector of the flakes is aligned poleward in response to the flow,
{
  since the radial component of the flow velocity is smaller than the other components along streamlines from the periphery to the pole near the stationary outer sphere.
}

{ Notably,} Fig.\ref{fig:virtual} shows the cross-section of the flow presented in Fig.\ref{fig:polar view}, where feather-like structures can be observed. Fig.\ref{fig:polar view} shows an anticlockwise extension from the base of the wings, whereas Fig.\ref{fig:virtual} illustrates a clockwise extension. This also indicates that the visualization results do not simply correspond to the flow structure.

\section{Discussion}

 Flow visualizations using flake particles as tracers have been performed in fluid dynamics experiments \cite{Van82,Sam03}.
  If the density ratio of the tracer to the working fluid is in unity and the Stokes number is sufficiently small, the distribution of tracers in the domain is expected to always be uniform from a theoretical viewpoint.
  Thus, the spatial fluctuation of the brightness observed experimentally was considered to originate not from the distribution but from the orientation of the particles.
  We primarily focused on the features of the flow field that can be read from the spatial pattern of the brightness in the spiral state of the SCF.
  In the following subsections, we discussed this in detail, especially with respect to SCF.

In \cite{Got11}, the authors conducted flow visualization using flake particles for the flow in a spherical vessel rotating under a small precession, which seems to be an almost rigidly rotating flow.
They reported that the spatial pattern of brightness visualized in their experiment originated not from the fluctuation of the number density of flakes but from the fluctuation of the orientation of flakes.
They successfully reproduced the experimentally observed pattern by numerically simulating the motion and rotation of flakes distributed in an almost rigid rotating flow. 
%

Notably, their numerical results imply that the bright regions in a visualized image are the superposition of those with flake orientation is isotropic and those with flakes tend to align in the direction in which the incident rays are reflected into the line of sight.
If the flakes are anisotropically oriented in a specific region, the incident light from a specific direction to that region is reflected only in a specific direction, and it is rare
for the reflecting rays to meet the optical axis of the camera.
In some partially observed regions, the direction of the reflected light matches the optical axis.
These regions are identified by the camera as strongly shrinking areas but only in limited areas.
By contrast, in the region where the flakes are isotropically oriented, 
 the incident light is equally reflected in all directions.
 Then, a part of the reflected light can be captured by a camera located in any direction.
{ In their experiment\cite{Got11}, most of the volumetric regions in the flow where flakes were well oriented must have been observed to be dark, except for a few exceptional areas (subtle bright sector-shaped pattern) where the direction of reflection matched the optical axis of the camera.
The regions that are bright due to flake isotropic orientation possess a planar structure in the bulk fluid, whereas the regions that are bright, owing to an orientation consistent with the optical axis, possess a volumetric structure.}

{
 Suppose a flake advecting through a flow field is traveling at a phase velocity. We assume that the time variation of the velocity gradient perceived by the flake is small from a Lagrangian perspective and the magnitude of vorticity is negligible compared to that of strain.
The Poincare–-Benedixon theorem\cite{Guk83,Hal91} guarantees the existence of limit cycles in a dynamic system in a two-dimensional closed domain if no fixed point exists.
The normal vector of the flake moves only in the unit sphere; thus, the theorem may be directly applied to the present stationary problems.
As the eigenvector of $\bfv{\nabla}\bfv{u}$ is constant in this case and the equation for $\bfv{s}$ is essentially linear, we can deduce that the orientation of the flake approaches the direction of the eigenvector associated with the smallest strain eigenvalue.
}

  The time constant required for the orientation to align may be estimated as the inverse of the absolute of the smallest (negative) real component among the eigenvalues of the velocity gradient tensor, {\rm that is }, $ 1/t_{\rm align} \sim |{\rm min}({\rm E.V. of } \bfv{\nabla u})|$.
  If the eigenvalues include a few complex conjugates, the argument below holds only when the remaining eigenvalue is negative.
  In a region with multiple fine-scale eddies, the orientation of the favorable eigenvector may significantly change within a short distance for which { flakes} advect during the interval required for the orientation of the flakes to align.
  Suppose that the eigenvector of the instantaneous velocity gradient favorable by a flake at $\bfv{x}$ { is represented by $\bfv{e}_1(\bfv{x})$ and that the flake moves from $\bfv{x}$ to $\bfv{x}'=\bfv{x}+\int_0^{t_{\rm align}} \bfv{u}(\bfv{x}(t),t) dt$ in $t_{\rm align}$}. 
 { In the region where the condition $|\bfv{e}_1(\bfv{x})\cdot \bfv{e}_1(\bfv{x}')|<1$ is realized, the orientations of the flakes are kept insufficiently aligned along the short path. }
  By contrast, for our spiral state of the SCF, flakes drift along spatially smooth but undulated streamlines, where $\bfv{e}(\bfv{x})\cdot \bfv{e}(\bfv{x}')\sim 1$ may be realized in parts of the entire domain.
  The 4-fold spiral state rotates at a slow phase speed around the rotation axis, and { the history experienced by the flakes along the trajectories, which crosses from a vortical structure to its neighboring structure in the spiral state, allows them to align in the preferred direction.}


According to Fig.\ref{fig:virtual}, the present spiral state of the SCF includes some regions where the brightness contour appears to be { gritty}. 
In such regions, the flakes are not oriented in a specific direction.
However, the background gradation, which is behind the { gritty pattern} associated with isotropic distribution, varies smoothly { from white to black} in space.
Such a pattern of light and dark suggests that the flake particles are well aligned.
Furthermore, this implies that the 4-fold arm-like shadow patterns obtained in the visualization experiments by Egbers\& Rath\cite{Egb95} were not the regions where the flake orientation remained isotropically distributed, as identified by the three sharp circles by Goto et al\cite{Got11}.

  \begin{figure}[ht]
    \centering
    \includegraphics[angle=0,width=0.78\columnwidth]{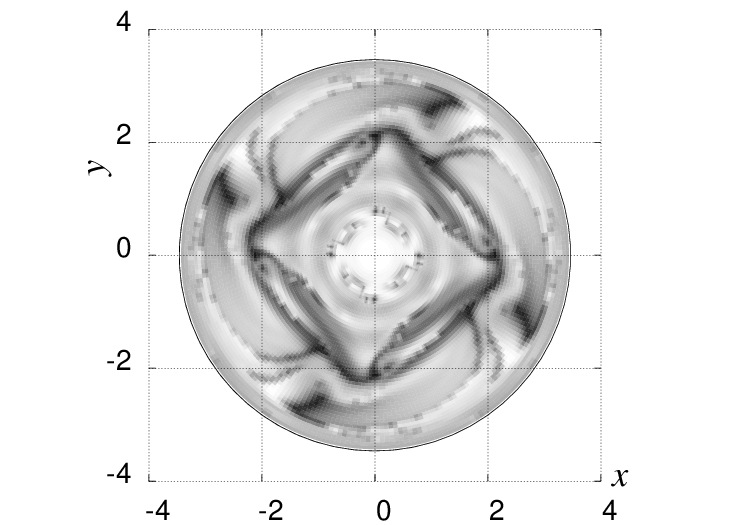}
    \caption{
      Similar to Fig.\ref{fig:virtual}, all initial directions of flakes were set constant ($\theta=0$).
    }
    \label{fig:oriented}
  \end{figure}

The same procedure shown in Fig.\ref{fig:virtual}(a) was carried out with initially distributing orientations of the flakes as a constant, $\bfv{s}=\bfv{e}_z$, as illustrated in Fig.\ref{fig:oriented}.
A comparison between Figs.\ref{fig:virtual}(a) and \ref{fig:oriented} indicates several flow characteristics in the present 4-fold spiral state.
First, the pattern of smoothly varying background brightness is consistent in both figures.
This suggests that a sufficiently long time had passed for the initial orientation to not { largely} affect the visualization results.
Both figures show that the orientation of the flakes {generating the background} was sufficiently acclimatized to the 4-fold state after the flakes experienced the shear of the spiral state.
Second, a { gritty pattern} associated with the isotropic distribution of the flake orientation remains, as shown in Fig.\ref{fig:virtual}. 
In the isotropic distribution, the orientation of the flakes is not fully aligned in some { regions, although} they are subjected to flow shear.
If the integral time is longer, some { regions in gritty}  may disappear and others may persist.
The degree of { gritty} depends on its position, and { the region in gritty}  possesses not a volumetric but planar structure, as shown in Fig.\ref{fig:oriented}.
Third, { gritty patterns} indicating isotropic distribution completely disappear around the axis (origin), thin layers near the outer spherical boundary, and the large black square.

  
  A common question that would arise is what brightness distribution represents in the flow field.
  As the flake particles traverse their orbits in a Lagrangian manner, they rotate under the effect of the shear from the flow field, which
  { varies in time.}
  In principle, predicting some characteristics of an instantaneous velocity field from a bright image is difficult,
  even if the flow is almost steady.
  { Considering this,}
  we investigated the correspondence between the brightness distribution drawn by the flakes and the instantaneous flow field.

  \begin{figure}[h]
    \centering
    \includegraphics[angle=0,width=0.78\columnwidth]{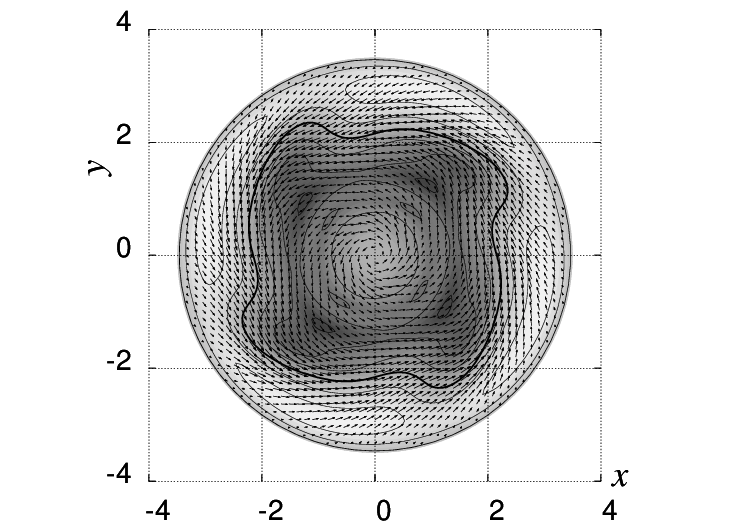}
    \caption{
      Velocity field of 4-fold state at $\Rey=600$ taken at $z=2$. Arrows indicate velocity fields $(u_x,u_y)$, and contour indicates the sign of $u_z$ ($u_z>0$ (white)).
    }
    \label{vector field at z=2}
  \end{figure}

Fig.\ref{vector field at z=2} is the projection of the vector of flow field at a cross-section $z=2$, where the shading and contours represent $u_z$ { and} the light colored areas correspond to $u_z>0$. 
A narrow square shadow tangential to the maximum diameter of the inner sphere is shown in Fig.\ref{fig:virtual}(a), exhibiting the brightness distribution at the cross-section $z=2$.
From a comparison with Fig.\ref{vector field at z=2}, no common feature is confirmed at the location of this square.
If we consider a windmill-like shape consisting of a field $u_z<0$, the slope against the $x$-axis is different from the square formed by the brightness distribution in Fig.\ref{fig:virtual}(a).
Apparently, velocity and brightness are not directly correlated at each point.

  \begin{figure}[ht]
    \centering
    \includegraphics[angle=0,width=0.78\columnwidth]{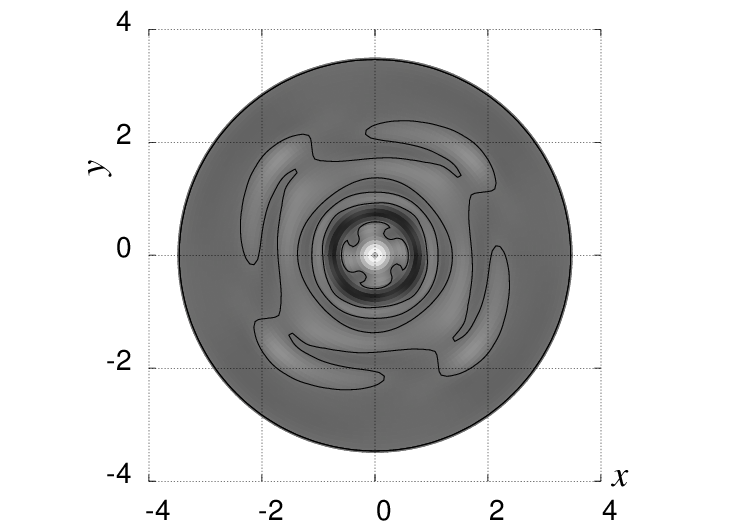}
    \caption{
      The Q criterion of the 4-fold state at $\Rey=600$ taken at $z=2$. Contour indicates the sign of Q ($Q>0$ (white)).
    }
    \label{Q at z=2}
  \end{figure}

A correlation between the brightness and velocity gradient is more probable.
The vorticity magnitude may be utilized to identify vortex structures; however, it does not always yield the desired coherent structures.
We focused on the Q criterion\cite{Hun88}, which is a commonly used method for vortex visualization, to identify the vortices in a flow with a positive second invariant of the velocity gradient tensor. 
Fig.\ref{Q at z=2} shows the flow field characterized in terms of the Q criterion, where the light colored area surrounded by the curves $Q=0$ are expected to possess a vortical structure.
The shearing of the stretch region between neighboring vortical structures may align the orientation of the flakes.
Because our spiral state does not contain sharp or fine eddy structures as in the turbulent flows 
 but only vague shear fluctuations extending from the poles to the equatorial plane in a spiral manner, the characterization of the flow field based on Q values is not evident.
We observed no significant correlation between the luminance distribution and 
{
  Q values, but a correlation 
}
 is evident from the comparison of Fig.\ref{Q at z=2} to the velocity field presented in Fig.\ref{vector field at z=2}. The velocity field and vortical structures estimated using the Q value are correlated.



{ Neither the velocity field nor the Q-value} could reproduce the luminance distribution produced by the flakes.
The present example is a counterexample, that is, for a general flow field, the brightness distribution of flakes obtained in the visualization experiments may
{ represent} 
neither the vortex structure nor the region of stretching at that instance,
{
  even if the intensity of the scattered light reflected from flakes is very sensitive to small velocity gradients\cite{Bel78}.
}.
Because we
{ could not } 
correlate it with the instantaneous field, we proposed a physical quantity that considers history.

  \begin{figure}[ht]
    \centering
    \includegraphics[angle=0,width=0.78\columnwidth]{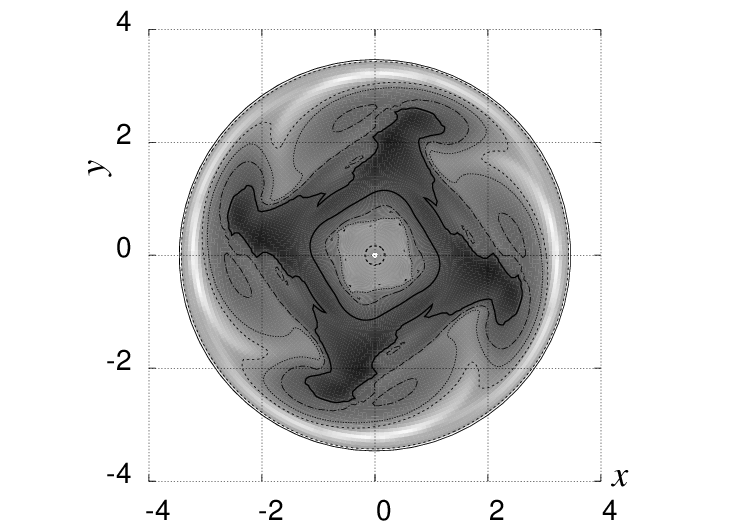}
    \caption{
      Shearing $\log{p}$ of the flow considering the background. Flakes in the white regions were advected with a relatively strong shear.
    }
    \label{histerisis}
  \end{figure}

{ The flakes are sufficiently small and stick to the flow to perform translational and rotational motions, so that the orientation of the flakes is strongly correlated with the stretching of the
fluid  elements.}
Taking this into account,
{ suppose a flake was present at $\bfv{x}=\bfv{x}(t)$ at a time $t$, then the following positive quantity is considered, $p^2(\bfv{x}) = \frac{1}{T}\int_0^{-T} \sum_{k=1,2,3} \lambda_k^2(\bfv{x}(t), t) dt$, where $\lambda_k$ is the eigenvalue of strain tensor at $\bfv{x}(t)$.
  { The contour map of $\log{p}$ in Fig.\ref{histerisis}} provides a pattern similar in Fig.\ref{fig:virtual},albeit with differences in the magnitude of the values.}
The value of $p$ takes on large values near the inner and outer spheres, which correlated with a pattern with the regions of background gradation varying smoothly in space, as seen in Fig.\ref{fig:virtual}.

\section{Summary}
The Newton--Raphson method was used to numerically { solve} the spiral state in a spherical Couette flow with an aspect ratio of $\eta=1/2$.
By { simulating} the translational and rotational motions of an infinitesimally small neutrally floating flaky particle with negligible inertia moving through the obtained state, an image obtained in the experimental visualization of the spiral state using thin aluminum flakes according to the experiment performed by Egbers and Rath [Acta Mech. 111 pp. 125--140 (1995)] { was} reproduced.
The numerically reproduced image agreed with the experimentally visualized images.
Several instantaneous physical variables calculated from the flow field of the spiral state could not reproduce the light and dark patterns owing to the flakes obtained during visualization.
Thus, incorporating the effects of the advection history of the flakes in the spiral state is necessary.

{
  The flicker contrast of the reflected light might increase with increasing Reynolds number, even for spiral states with the same wavenumber. Also, as the degree of flake orientation alignment increases, the intensity of the reflections will change accordingly. This means that the intensity of flow shear may be quantitatively estimated from the visualization. Furthermore, the degree of flake alignment may fall rapidly as the flow transitions to turbulence consisting of small eddies, and the distance and time scale of the turbulence may be estimated. These points will be pursued in future studies.
}

\begin{acknowledgments}
  The authors would like to thank Mr. F. Goto and Mr. T. Inagaki for their pilot surveys and Mr. Yoshikawa for his valuable comments on the draft.
{ They also acknowledge ORDIST in Kansai University
    and the RIMS Joint Research Activities in Kyoto University for providing a space for their research and communication.}
This work was supported in part by a Grant-in-Aid for Scientific Research(C) and JSPS KAKENHI Grant Nos.20K04294 and 24K07331.
This study also benefited from the interaction within RISE-2018 No.824022 ATM2BT of the European Union Horizon 2020-MSCA program, which includes Kansai University.
\end{acknowledgments}

\bibliography{scf12}

\end{document}